\documentclass[11pt,a4paper]{article}
\usepackage[utf8]{inputenc}

\usepackage{framed}
\usepackage{mdframed}

\usepackage{fullpage}
\usepackage[T1]{fontenc}

\usepackage{epsf}
\usepackage{graphicx}
\usepackage{verbatim}
\usepackage{color}	
\usepackage{bbold}
\usepackage{hyperref}
\usepackage[dvipsnames]{xcolor}
\hypersetup{
     colorlinks=true,
     linkcolor = BrickRed,
     anchorcolor = blue,
     citecolor = PineGreen,
     filecolor = blue,
     urlcolor = blue
     }

\usepackage[most]{tcolorbox}
\usepackage{empheq}
\newtcbox{\mymath}[1][]{%
    nobeforeafter, math upper, tcbox raise base,
    enhanced, colframe=blue!30!black,
    colback=blue!30, boxrule=1pt,
    #1}

\usepackage[british]{babel}
\usepackage{verbatim}
\usepackage[T1]{fontenc}
\usepackage{lmodern}
\usepackage{lipsum}
\usepackage{booktabs}
\usepackage{caption}
\usepackage{cite}
\usepackage{soul,color}
\usepackage[toc,page]{appendix}

\usepackage{ifpdf}
\usepackage{cancel}

\usepackage{amsthm, amssymb}
\usepackage{pifont}
\usepackage{bm}
\usepackage{mathtools}
\usepackage{amstext}
\usepackage{braket}
\usepackage{multirow}

\usepackage[normalem]{ulem}
\usepackage{xcolor}
\usepackage{cancel}

\newcommand\redsout{\bgroup\markoverwith{\textcolor{red}{\rule[0.5ex]{2pt}{0.4pt}}}\ULon}

\begin{document}
\vspace{5mm}
\vspace{0.5cm}

\def\be{\begin{eqnarray}}
\def\ee{\end{eqnarray}}

\def\ba{\begin{aligned}}
\def\ea{\end{aligned}}

\def\ls{\left[}
\def\rs{\right]}
\def\lc{\left\{}
\def\rc{\right\}}

\def\p{\partial}

\def\S{\Sigma}

\def\s{\sigma}

\def\O{\Omega}

\def\a{\alpha}
\def\b{\beta}
\def\g{\gamma}

\def\ad{{\dot \alpha}}
\def\bd{{\dot \beta}}
\def\gd{{\dot \gamma}}
\newcommand{\ft}[2]{{\textstyle\frac{#1}{#2}}}
\def\ib{{\overline \imath}}
\def\jb{{\overline \jmath}}
\def\Re{\mathop{\rm Re}\nolimits}
\def\Im{\mathop{\rm Im}\nolimits}
\def\trace{\mathop{\rm Tr}\nolimits}
\def\rmi{{ i}}

\def\N{\mathcal{N}}

\newcommand{\SU}{\mathop{\rm SU}}
\newcommand{\SO}{\mathop{\rm SO}}
\newcommand{\U}{\mathop{\rm {}U}}
\newcommand{\USp}{\mathop{\rm {}USp}}
\newcommand{\OSp}{\mathop{\rm {}OSp}}
\newcommand{\Symp}{\mathop{\rm {}Sp}}
\newcommand{\Sl}{\mathop{\rm {}S}\ell }
\newcommand{\Gl}{\mathop{\rm {}G}\ell }
\newcommand{\Spin}{\mathop{\rm {}Spin}}
\newcommand*\mybluebox[1]{\colorbox{blue!20}{\hspace{1em}#1\hspace{1em}}}

\newlength{\Lnote}
\newcommand{\notte}[1]
     {\addtolength{\leftmargini}{4em}
    \settowidth{\Lnote}{\textbf{Note:~}}
    \begin{quote}
    \rule{\dimexpr\textwidth-2\leftmargini}{1pt}\\
    \mbox{}\hspace{-\Lnote}\textbf{Note:~}%
    #1\\[-0.5ex] 
    \rule{\dimexpr\textwidth-2\leftmargini}{1pt}
    \end{quote}
    \addtolength{\leftmargini}{-4em}}
\def\hc{c.c.}

\numberwithin{equation}{section}

\allowdisplaybreaks

\allowbreak


\hfill\text{LAPTH-052/23}

\thispagestyle{empty}
\begin{flushright}

\end{flushright}
\vspace{35pt}

\begin{center}
	    {  \bf{\LARGE{
	    Anisotropic scale-separated AdS$_4$ flux vacua 
        \vspace{0.1cm} 
	    }}}

		\vspace{50pt}

    	{\Large George~Tringas}

		\vspace{30pt}

{

{ \large Laboratoire d'Annecy-le-Vieux de Physique Théorique {(LAPTh)},\\
CNRS, Université Savoie Mont Blanc (USMB), UMR 5108, \\
9 Chemin de Bellevue, 74940 Annecy, France }

}


\vspace{1.3cm}

E-mails: tringas@lapth.cnrs.fr

\vspace{1.3cm}

{ABSTRACT}

\end{center}

We present minimally supersymmetric AdS$_4$ flux vacua derived from massive type IIA compactified on $T^6/\mathbb{Z}_3\times \mathbb{Z}_3$ orbifold, characterized by unconstrained fluxes with general scaling.
We discover anisotropic scaling solutions where scale separation is realized in the supergravity limit and the subvolumes of the internal space become large and anisotropic for large values of unconstrained fluxes.
Additionally, we identify regimes where subvolumes are either shrinking or remain constant while scale separation is either broken or realized.
Then we employ a probe anti-D4-brane to interpolate between vacua, finding that it interpolates through the regimes we previously identified.
Finally, we utilize an open string modulus of the anti-D4-brane to calculate the distance between vacua for the regime where scale separation is realized in the supergravity limit.
We show the dependence of both the geodesic distance and the distance conjecture parameter on the flux scaling.

\thispagestyle{empty} 
\setcounter{page}{0}

\baselineskip 6mm

\newpage


{\hypersetup{hidelinks}
\tableofcontents
}

\setcounter{footnote}{0}

\pagebreak

\section{Introduction}

String theory is considered one of the prominent theories of quantum gravity, however the derivation of solutions which can describe the current state of our universe seem to be non-trivial. Since critical string theory requires the existence of ten dimensions, while we observe only four, one of the central requirements for achieving a realistic phenomenological scenario is to hide the extra dimensions. One way to get this is to assume the extra dimensions to be compact and periodical while the higher dimensional part of the fields is described by the Kaluza-Klein (KK) modes which indicate their presence. The qualitative condition that we use in order to see whether there is a large energy gap between extra dimensional states of the fields and the vacuum energy of the effective theory, is to compare the curvature radius of external space to the scale of the Kaluza-Klein modes
\begin{align}
    \frac{L_{\text{KK}}}{L_{\Lambda}} \ll 1\,.
\end{align}

An essential feature of the effective theory, in order to explain the positive vacuum energy observed in our universe, is the search for stable and scale-separated (quasi)-de Sitter solutions, however it turns out to be a challenging problem \cite{Danielsson:2018ztv}-\cite{Garg:2018reu}. To construct such vacua within string theory, one can consider incorporating quantum effects \cite{Kachru:2003aw}-\cite{Conlon:2005ki} (taking into account recent criticisms \cite{Sethi:2017phn}-\cite{Junghans:2022exo}), or investigate proposals in more general frameworks or with more ingredients \cite{Antoniadis:2018hqy}-\cite{Bena:2022cwb}. However, the most economical approach is to construct {\it classical} de-Sitter solutions from string theory, but this presents difficulties too \cite{Andriot:2020wpp}-\cite{Andriot:2022bnb}. From the perspective of lower-dimensional supergravity, which involves the aspect of supersymmetry breaking, obtaining such constructions is also a challenging problem, see \cite{Cribiori:2020use}-\cite{Farakos:2022jcl}.

We now turn our attention to effective theories within AdS vacua and address a question analogous to that in the case of de Sitter: Can we discover scale-separated AdS solutions within string theory? In four dimensions minimally supersymmetric AdS$_4$ vacua from type IIA where analytically constructed in \cite{DeWolfe:2005uu}, and then in \cite{Camara:2005dc}, while the connection of scale-separated AdS$_4$ from type IIA to M-theory was discussed in \cite{Cribiori:2021djm}. 
Vacua with scale separation in the same setup, where the radii are small compared to the string length, and the focus is on the study of T-dualities, were recently investigated in \cite{Carrasco:2023hta}.
For effective theories with less dimensions coming form type IIA, scale separation can be achieved in minimally supersymmetric AdS$_3$ \cite{Farakos:2020phe}-\cite{Farakos:2023nms}, however in two dimensions it is still an open problem \cite{Lust:2020npd}. In type IIB attempts have been made but it is not clear whether it is possible to achieve it \cite{Petrini:2013ika}-\cite{Emelin:2020buq}, while constructions involving modest scale separation have been examined in \cite{Tsimpis:2022orc}. A common and necessary ingredient of those setups are the smeared orientifolds \cite{Gautason:2015tig}, while their backreaction and the validity of the approximation have been studied in \cite{Blaback:2010sj}-\cite{Baines:2020dmu} and using an expansion procedure in \cite{Junghans:2020acz}-\cite{Junghans:2023lpo}. 
The connection between scale-separated AdS vacua and their corresponding CFTs has been studied in the context of dimensions of conformal operators and central charges in \cite{Conlon:2021cjk}-\cite{Plauschinn:2022ztd} while bounds on the masses and possible modification are studied in \cite{Andriot:2022brg}-\cite{AndriotTringas}.
On the other hand, challenges associated with constructing scale-separated vacua are examined in \cite{Tsimpis:2012tu}-\cite{Font:2019uva}, while achieving scale separation from the perspective of lower-dimensional supergravity with supersymmetry $N\geq 2$ does not seem to be possible \cite{Cribiori:2022trc}, \cite{Cribiori:2023gcy},\cite{Green:2007zzb}.

The possibility of discovering consistent supersymmetric scale-separated AdS vacua has been both disfavored and conjectured. This consideration involves measuring the geodesic distance and the KK modes for the full field space metric, as discussed in \cite{Lust:2019zwm}. An earlier work, \cite{Ooguri:2006in}, known as the swampland distance conjecture, focuses solely on the scalar field space and places restrictions on the behavior of KK masses when traveling into the moduli space. See also discussion on the distance and de-Sitter \cite{Ooguri:2018wrx}.
However, despite these conjectures, there are counterexamples from 10d classical supergravity that challenge such propositions. For reviews on the swampland, compactifications, and related topics discussed previously, see \cite{Palti:2019pca}-\cite{VanRiet:2023pnx}.


In this paper, our focus is on supersymmetric AdS$_4$ vacua derived from the compactification of massive type IIA on the orbifolded torus $T^6/\mathbb{Z}_3\times\mathbb{Z}_3$. Vacua of such configurations were examined in \cite{DeWolfe:2005uu} (so called DGKT), where it was determined that they incorporate all the elements we previously discussed.
An important feature of \cite{DeWolfe:2005uu} is that the sizes of the volumes, the $g_s$ coupling, and scale separation are parametrically governed by the unconstrained flux of field strength $F_4$, which is expanded in the internal space, fills three different four-cycles and its flux components $F_4^{(i)}$, with $i=1,2,3$ , scale the same, as $F_4^{(i)}\sim N$.
For large values of the flux, $N\gg 1$, the curvature radius of external space and the Kaluza-Klein scale decouple, while the volume becomes large and the string coupling small.
The two latter features describe a \textit{classical solution}, where we can treat string theory perturbatively, ignoring strong coupling effects while also suppressing $\alpha^{\prime}$-corrections.
These are satisfied by qualitatively imposing
\begin{align}
    R_i\gg l_s\,,\quad\quad
    \text{vol}\gg l_s^{\,d}\,,\quad\quad
    g_s< 1\,,
\end{align}
where the first two quantities correspond to the radii and the volume of the internal space respectively, $d$ stands for the dimensions of the internal space, $l_s$ is the string length and $g_s\equiv \langle e^{\phi}\rangle$ the string coupling.

The novel aspect of our paper is that, through an examination of the scaling of the dimensionally scalar reduced potential, we keep arbitrary distinct scalings $f_i$ with $i=1,2,3$ for all three components of the $F_4^{(i)}\sim N^{f_i}$ fluxes, thereby uncovering a new general scaling freedom.
The $g_s$ coupling, the volumes and the scale separation condition are again determined by the unconstrained flux of $F_4$, but their scaling behavior now also depends on the general scalings $f_i$. 
Due to this freedom in scaling, the subvolumes are allowed to scale differently, resulting in anisotropy, as previously observed in \cite{Farakos:2023nms}.
We discover four new general families (or regimes) of solutions, determined by inequalities among the three scalings $f_i$, characterized by whether the solutions are in the supergravity regime and whether scale separation is realized.
For instance, for large flux values and specific values of the scalings, scale separation may be realized in the supergravity regime with either anisotropic or isotropic large radii, while for another scaling ansatz, scale separation may be realized with radii that are either constant and small or shrinking. Additionally broken scale separation with shrinking radii is one of the solutions within these families, although our primary focus and emphasis are on highlighting the presence of scale separation with anisotropic radii in the supergravity limit.

Then, we employ a space filling anti-D4-brane, which is codimension-one in the six directions transverse to one of the four-cycles related to one component of $F_4$, in order to interpolate between vacua, and we find that it interpolates through all the new vacua discovered.
We discover that within this landscape of vacua, there exist anisotropic, scale-separated vacua in the supergravity limit, surrounded by scale-separated or non-scale-separated vacua with small subvolumes, breaking the validity of the classical solution.
Finally, as an application, we measure the distance between the vacua located at the boundaries of the region containing scale-separated classical vacua and observe a dependence of both the geodesic distance and the distance conjecture parameter on the change in flux scaling.

\section{AdS$_4$ from massive type IIA}\label{section2}

In this section, we present the setup of massive type IIA supergravity, our conventions, and provide a brief overview of the internal space of our compactification. All of these elements can be collectively found in \cite{DeWolfe:2005uu}. Here, we use the basics of these ingredients, which allow us to derive the four-dimensional potential and determine the general scaling for the fluxes. For a more detailed discussion of \cite{DeWolfe:2005uu} including warping, see \cite{AndriotTringas}.

\subsection{The setup: actions, internal space and equations of motion}
The ten-dimensional action of massive type IIA supergravity written in the string frame is
\begin{align}
S_{\text{IIA}}&=\frac{1}{2\kappa^2_{10}}\int \text{d}^{10}X\sqrt{-G}\left( e^{-2\phi}\left(R_{10}+4\partial_M\phi\partial^M\phi-\frac{1}{2}\vert H_3\vert^2\right)-\frac{1}{2}\sum_p\vert F_{p}\vert^2\right) \,,  \label{ActionTypeIIA}
\end{align}
where $X^{M}$ denotes the ten-dimensional coordinates and $G\equiv\text{det}[G_{MN}]$ the ten-dimensional metric. For the field strengths we have $\vert F_p\vert^2=\frac{1}{p!}F_{M_1\dots M_p}F^{M_1\dots M_p}$ with $p=0,2,4$. The action of the relevant orientifold plane for our analysis is given by
\begin{align}
    S_{\text{O6}}&=\mu_{\text{O6}}\int\text{d}^{7}\xi e^{-\phi}\sqrt{-g_7}
    +\mu_{\text{O6}}\int C_7 \,,
\end{align}
where $\xi^i$, with $i=1,\dots,7$, stand for the worldvolume coordinates. The O6-planes here are space filling, they fill the four-dimensional external spacetime, and wrap four different three-cycles, $\alpha_i$ with $i=1,\dots 4$, of the internal space. The conventions for the scales and the tensions of local objects that we use are the following
\begin{align}
    2\kappa_{10}^2=(2\pi)^7\alpha^{\prime 4} \,,\quad\quad
    \alpha^{\prime}=l_s^2
    \,,\quad\quad 
    \mu_{\text{Dp}}=(2\pi)^{-p}(\alpha^{\prime})^{-\frac{p+1}{2}}\,,\quad\quad 
    \mu_{\text{Op}}=-2^{p-5}\times \mu_{\text{Dp}} \,.
\end{align}
We will be interested in a flux background where the external space is AdS$_4$ and the internal space is compact and we make an ansatz for the ten-dimensional metric of the form
\begin{align}
    \text{d}s^2_{10}=g_{\mu\nu}\text{d}x^{\mu}\text{d}x^{\nu}+g_{mn}\text{d}y^{m}\text{d}y^{n} \,,
\end{align}
where $g_{\mu\nu}$ is the unwarped four-dimensional external metric, $g_{mn}$ is six-dimensional internal one while $x^{\mu}$ and $y^m$ are the external and internal coordinates respectively. Since the internal space is $T^6=T^2\times T^2\times T^2$ the sizes, or deformations, of the $T^2$s can be described as fluctuations in the four-dimensional effective theory. The internal space metric can be parametrized in the following way
\begin{align}\label{intermetric}
    \text{d}s^2_{6}=2(\sqrt{3})^{1/3}\sum_{i=1}^3\upsilon_i\,\left((\text{d}y^{2i-1})^2+(\text{d}y^{2i})^2\right) \,,
\end{align}
where we just have written in cartesian coordinates the product of $\text{d}z^i\text{d}\overline{z}^i$ where $z^i=y^{2i-1}+iy^{2i}$ with $i=1,2,3$ . The moduli $\upsilon_i$ describe the deformations of $T^2s$ with the metric dependence $g_{mn}=2(\sqrt{3})^{1/3}\,\upsilon_{i}\tilde{g}_{mn}$ and have only spacetime dependence $\upsilon_i\equiv\upsilon_i(x)$.
The overall constant arises from the specific normalization of the integrated volumes, while the presence of two $\mathbb{Z}_3$ symmetries constructs the orbifold $T^6/\mathbb{Z}_3\times\mathbb{Z}_3$, which is a singular limit of a Calabi-Yau. For details e.g. on the symmetry actions, see \cite{DeWolfe:2005uu}.
The internal volume\footnote{Note that $\int_{T^6/\mathbb{Z}_3^2}\text{d}^6y=\frac{1}{8\sqrt{3}}$.} is defined as
\begin{align}
    \text{vol}=\int \text{d}^6y\sqrt{g_6}= \upsilon_1\upsilon_2\upsilon_3 \,.
\end{align}
For completeness, it is important to also present the forms that are invariant under the $\mathbb{Z}_3$ symmetries, as fluxes are expanded over them and local sources wrap around them. The invariant two-form, and its Hodge dual, can be expanded on the basis
\begin{align}
    w^i=(\sqrt{3})^{1/3}i\text{d}z^i\wedge\text{d}\overline{z}^i \,,
    \quad\quad \tilde{w}^i= w^j\wedge w^k \,,
\end{align}
with $i,j,k=1,2,3$. Three-forms which are invariant under the symmetries of the orbifold can be deduced in the following way
\begin{align}
    \Omega=3^{1/4}i\text{d}z^1\wedge\text{d}z^2\wedge\text{d}z^3=\frac{1}{\sqrt{2}}(\alpha+i\beta) \,.
\end{align}
The basis $\alpha$ consists of four terms $\alpha=\sum_{i=1}^4\alpha_i$, which are the cycles wrapped by the O6-planes, while $\beta$ consists of the directions $\beta=\sum_{i=1}^4\beta_i$ where the O6-planes are localized at.

Now, we focus on the equations of motion derived from the 10d supergravity action that are relevant to our analysis. We use the smeared approximation where the dilaton is assumed to be slowly varying with respect to the internal coordinates $\phi(y)\approx\phi$, the internal space is Ricci flat ($R_{mn}=0$), and the background field strengths are expanded in terms of harmonic cycles of the internal space 
\begin{align}\label{dF4}
    \text{d}F_p=\text{d}\star F_p=0\,,\quad\quad
    \text{d}\phi=0 \,.
\end{align}
The Bianchi identities are satisfied due to the presence of O6-planes and for proper choice of $F_2=0$ we get 
\begin{align}\label{SmearedBianchi}
    \text{d}F_2=0=H_3\wedge F_0-(2\pi)^7\alpha^{\prime 4}\mu_{\text{O6}}\sum j_{\beta_i}\,,\,\quad\quad
    \text{d}F_4=0=H_3\wedge F_2\,.
\end{align}
In our analysis the $F_2$ flux is chosen to be zero but in such setups it can be non-zero and expanded over the harmonic two-form basis $w_a$ of the Calabi-Yau. The definition of the currents which are involved into the equations of motion is given by
\begin{align}
    j_{\beta_i}=\frac{\text{d}^{3}y\sqrt{g_{\beta_i}}}{\int_{\beta_i}\text{d}^3y\sqrt{g_{\beta_i}}} \,,
\end{align}
and $g_{\beta_i}$ represents the metrics of the cycles transverse to the relevant O6-plane associated with the basis $\beta$. The remaining form-field equations are
\begin{align}\label{SmearedBianchi2}
    0&=H_3\wedge\star_{6}F_4\,,\quad\quad
    0=H_3\wedge\star_{6}F_6\,,\quad\quad
    0=\star_{6}F_2\wedge F_0+\star_{6}F_4\wedge F_2+\star_{6}F_6\wedge F_4\,.
\end{align}
The Hodge dual $\star_6F_p$ is a harmonic form, and consequently the terms resulting from wedging with other forms to create five-forms are also harmonic forms which vanish on a Calabi-Yau. The terms involving $F_2$ vanish due to the flux ansatz, thus the equations of motion are satisfied. From the equations above, we observe that $F_4$ is not bounded, but only $H_3$ is constrained by the Bianchi identity. The equations \eqref{SmearedBianchi} and \eqref{SmearedBianchi2} tell us that all the forms are closed at the smeared limit.

\section{Scaling analysis}\label{section3}

In this section, we present a general flux scaling ansatz derived from the potential after dimensional reduction.
We express all relevant quantities, including fluxes, fields, and volumes, in terms of the scalings associated with $F_4$. Subsequently, we determine the KK scale to calculate the dependence of scale separation on these general scalings.

\subsection{Fluxes and effective scalar potential}

After introducing the internal space for compactification, we introduce the fluxes that are invariant under the orbifold and their basis related to the internal space
\begin{align}
    F_4=e_i\tilde{w}^i \,, \quad\quad H_3=-p\beta \,,\quad\quad F_0=m_0 \,,
\end{align}
as we have considered the simplest case leading to full moduli stabilization, with $F_2=0$. The quantities $\tilde{w}^i$ and $\beta$ are the invariant basis introduced before, so the fluxes $e_i$ and $p$ carry all the scaling dependence. It is essential to note that $F_4$ is expanded over three distinct internal four-cycles,\footnote{
    $\tilde{w}^1\sim\text{d}y^3\wedge\text{d}y^4\wedge\text{d}y^5\wedge\text{d}y^6\,,\,\,\,
    \tilde{w}^2\sim\text{d}y^1\wedge\text{d}y^2\wedge\text{d}y^5\wedge\text{d}y^6\,,\,\,\,
    \tilde{w}^3\sim\text{d}y^1\wedge\text{d}y^2\wedge\text{d}y^3\wedge\text{d}y^4$ }
such that there is one distinct flux filling each cycle
\begin{align}
    F_4=\sum_{i=1}^3F_4^{(i)}=e_1\tilde{w}^1+e_2\tilde{w}^2+e_3\tilde{w}^3 \,.
\end{align}
Now we turn to the four-dimensional effective scalar potential. Starting from the action \eqref{ActionTypeIIA}, we perform the dimensional reduction and the following Weyl rescaling 
\begin{align}
    g_{\mu\nu}=\frac{e^{2\phi}}{\text{vol}}\tilde{g}_{\mu\nu} \,,
\end{align}
where $\tilde{g}_{\mu\nu}$ is the four-dimensional Einstein metric, and we obtain the effective action in the Einstein frame, which schematically takes the following form
\begin{align}
    S=\frac{1}{2\kappa_{10}^2}\int \text{d}^4x\sqrt{-\tilde{g}_4}\left(M^2_{\text{Pl}}R_4+\dots - V\right) \,,
\end{align}
with $M_{\text{Pl}}^2$ being constant in this frame. Since we have first performed the dimensional reduction and then the Weyl rescaling, the potential is given by $V = \frac{e^{4\phi}}{\text{vol}} \times \tilde{V}$, where $\tilde{V}$ represents the field strengths expanded over the internal coordinates and the rescaled O6-plane contribution from the string frame type IIA action. The effective four-dimensional potential in this frame is as follows
\begin{align}\label{potential}
    V&=
    \frac{1}{2}\frac{e^{2\phi}}{\text{vol}^2}p^{2}
    +\frac{1}{2}\left(e_1^2\upsilon_1^2+e_2^2\upsilon_2^2+e_3^2\upsilon_3^2\right)\frac{e^{4\phi}}{\text{vol}^3}
    +\frac{m_0^2}{2}\frac{e^{4\phi}}{\text{vol}}
    -2m_0p \frac{e^{3\phi}}{\text{vol}^{3/2}} \,.
\end{align}
The potential consists of four terms: the first corresponds to $\vert H_3\vert^2$, the second to $\vert F_4\vert^2$, the third is the Romans mass, and the last one arises from the O6-plane action, obtained by exchanging the tension with fluxes using the Bianchi identity. It is important to note that the axions $B_2=\sum_{i=1}^3b_iw^i$ and $C_3=\xi\,\alpha$ have background values of their moduli set to $b_i=\xi=0$ since they appear quadratically in the potential, see \cite{DeWolfe:2005uu}.

Then, we require the fluxes to have the following scaling:
\begin{align}\label{fluxes}
    p\sim N^0\,,\quad e_1\sim N^{f_1}\,,\quad e_2\sim N^{f_2}\,,\quad e_3\sim N^{f_3}\,,\quad m_0\sim N^0
\end{align}
and for the moduli, dilaton and subvolumes 
\begin{align}
    e^{\phi}\sim N^d\,,
    \quad \upsilon_1\sim N^{u_1} \,,
    \quad \upsilon_2\sim N^{u_2} \,,
    \quad \upsilon_3\sim N^{u_3} \,.
\end{align}
Since $F_0$ is a quantized constant it does not scale, thus its quantum $m_0 \sim N^0$. Then, from the Bianchi identity in \eqref{SmearedBianchi} and since $j_{\beta_i,3}\sim N^0$ does not scale, the flux of $H_3$ is imposed not to scale $p\sim N^0$ as well.

Next, we require all the terms in the effective potential in \eqref{potential} to scale the same
\begin{align}
& 2d-2u_1-2u_2-2u_3=a  \,,
\label{scaleq1}\\
& 2f_1+2u_1+4d-3u_1-3u_2-3u_3=b  \,, \label{scaleq2}\\
& 2f_2+2u_2+4d-3u_1-3u_2-3u_3=c  \,, \label{scaleq31}\\
& 2f_3+2u_3+4d-3u_1-3u_2-3u_3=d  \,, \label{scaleq32}\\
& 0+4d-u_1-u_2-u_3=e  \,,\label{scaleq33}\\
& 0+3d-\frac{3}{2}u_1-\frac{3}{2}u_2-\frac{3}{2}u_3=g  \,,\label{scaleq44}
\end{align}
where the right-hand side of the equations just represents the overall scaling of each term; the equation \eqref{scaleq1} corresponds to the first term of the potential, the equations \eqref{scaleq2}-\eqref{scaleq32} correspond to the second term etc.
We require \eqref{scaleq1}-\eqref{scaleq44} to scale the same
\begin{align}
    a=b=c=d=e=g \,,
\end{align}
and from these equations we find the general scaling for the subvolumes
\begin{align}\label{subvolumes}
    \upsilon_1\sim N^{\frac{1}{2}(-f_1+f_2+f_3)}\,,\quad\quad
    \upsilon_2\sim N^{\frac{1}{2}(f_1-f_2+f_3)}\,,\quad\quad
    \upsilon_3\sim N^{\frac{1}{2}(f_1+f_2-f_3)}\,, 
\end{align}
while for the dilaton and the overall internal space volume 
\begin{align}\label{dilvol}
    e^{\phi}&\sim N^{-\frac{1}{4}(f_1+f_2+f_3)}\,,\quad\quad
    \text{vol}\sim N^{\frac{1}{2}(f_1+f_2+f_3)} \,.
\end{align}
Notice that the scalings of all the quantities have been expressed in terms of the scaling of the three components of $F_4$.
One can derive the scaling mentioned above from the ten-dimensional equations of motion (6d, 4d Einstein and dilaton equations). Alternatively, one can verify that the 10d equations of motion remain invariant under the general scaling of fields and fluxes that we have found. 
It is important to note that for $\sum_{i=1}^3f_i>0$, the solution will always be in the large volume and small coupling regime. However, in our analysis, we will also consider zero scalings and also discuss on negative scalings and their implications.
The scaling of the potential $V$ and, consequently, the vacuum expectation value $\langle V \rangle$ scales as follows:
\begin{align}
    V\sim\langle V\rangle=\frac{\Lambda_4}{M^2_{\text{Pl}}}\sim N^{-\frac{3}{2}(f_1+f_2+f_3)} \,,
\end{align}
while for completeness we express it in terms of the cosmological constant $\Lambda_4$ and the Planck mass, where the latter does not scale $M_{\text{Pl}}^2\sim N^0$ in the Einstein frame.

Whether the vacua of these models preserve supersymmetry depends on the signs of the $e_i$ fluxes.
At the vacuum, the following conditions should be satisfied: $\text{sgn}(m_0e_1e_2e_3) < 0$, and $e_1\upsilon_1 = e_2\upsilon_2 = e_3\upsilon_3$, which requires all the signs of $e_i$ to be the same and is not related to the scaling of the fluxes.
One can see that the latter condition is always satisfied by \eqref{fluxes} and \eqref{subvolumes}.
Thus, the general scaling always satisfies the supersymmetric vacuum conditions, and as a result, the moduli will always be stabilized.

\subsection{Scale separation}

As we can observe from the previously derived scaling, anisotropy in the fluxes (different values for $f_i$) leads to anisotropy in the subvolumes of the internal space, or vice versa. Consequently, since the KK scale is analogous to the radii of the internal space, one cannot assume that there will be a single KK scale; instead, there will be different KK scales due to the anisotropic radii. In this subsection, we provide a more analytical treatment and demonstrate that there are multiple conditions to consider for determining whether scale separation is broken or not. These conditions involve comparing the KK scale to the AdS scale or equivalently comparing the KK masses to the vacuum energy.

While potentially all the moduli have KK modes, we will specifically focus on the KK modes of the dilaton and examine how their masses are influenced by the general scaling of the subvolumes. In the case of a toroidal internal space, we identify the periodicity of the internal space coordinates as $y^m\simeq y^m+1$. Due to dimensional reduction, a field $\phi(X^M)$ is also periodically identified as $\phi(x^{\mu},y^{m})\equiv\phi(x^{\mu},y^{m}+1)$.
We decompose the field in Fourier modes on the circle
\begin{align}
    \phi(x^{\mu},y^m)=\sum_{k=0}^{\infty}\phi_k(x^{\mu})\,\text{cos}[2\pi k\,y^m] \,,
\end{align}
where, $\phi_k(x^\mu)$ represents the four-dimensional scalar fields. The mode $\phi_0(x^\mu)$ stands for the standard (zero mode) dilaton kinetic term, while the others make up the tower of modes. For our purposes, we only need to calculate the first mode of the tower, using the ansatz
\begin{align}
    \phi(x^{\mu},y^m)=\phi_0(x^{\mu})+\phi_1(x^{\mu})\,\text{cos}[2\pi k\,y^m] \,.
\end{align}
In order to obtain the KK masses, we perform the Weyl rescaling, use the ansatz above for the scalar field and perform the dimensional reduction to get
\begin{equation}
\begin{split}
    S\sim &\int \text{d}^4x\sqrt{-\tilde{g}_4}\int \text{d}^6y\,\Big(\tilde{R}_4
    -\frac{1}{2}\tilde{g}^{\mu\nu}\partial_{\mu}\phi_0\partial_{\nu}\phi_0+ \\
    &-\frac{1}{2}(\text{cos}[2\pi y^m])^2\tilde{g}^{\mu\nu}\partial_{\mu}\phi_1\partial_{\nu}\phi_1
    -\frac{1}{2}(2\pi\,\text{sin}[2\pi y^m])^2\sum_{i=1}^3\frac{1}{\upsilon_i}\frac{e^{2\phi}}{\text{vol}}\,\phi_1^2
    +\dots \Big) \,.
\end{split}
\end{equation}
We have extracted the moduli dependence from the internal space metric, denoted as $g_{mn} \sim \upsilon_{i} \tilde{g}_{mn}$, with $\tilde{g}_4 = \text{det}[\tilde{g}_{\mu\nu}]$. We use the approximation symbol due to the overall numerical factors that we have omitted. Now, this allows us to read off the first KK masses, which are
\begin{align}
    m^2_{\text{KK}_i}\sim\frac{e^{2\phi}\text{vol}^{-1}}{\upsilon_i}
    \sim N^{-\frac{1}{2}f_i-\frac{3}{2}(f_j+f_k)} \,,
\end{align}
where $i,j,k=1,2,3$. For isotropic scaling, e.g., $f_i=1$, we recover the KK mass scaling $m^2_{\text{KK}}\sim M^2_{\text{Pl}}/L^2_{\text{KK}}\sim N^{-7/2}$ as found in \cite{DeWolfe:2005uu}. Again, in the Einstein frame\footnote{We also express all the relevant quantities in the string frame. 
The Planck mass in this frame is dynamical and scales as $M^2_{\text{Pl}}=\text{vol}\,e^{-2\phi}\sim N^{f_1+f_2+f_3}$. We observe that the sum of the exponents cannot be negative or zero, otherwise the Planck mass vanishes or shrinks. 
We introduce the cosmological constant $\Lambda_4$, defined as the minimum of $\tilde{V}$, where the potential $\tilde{V}$ arises directly from the fluxes and sources in the string frame action, multiplied by $e^{-2\phi}$ and scales as $\Lambda_4\equiv\langle\tilde{V}\rangle\sim N^{-\frac{1}{2}(f_1+f_2+f_3)}$.
We provide the scaling of the AdS radius $L_{\text{AdS}}\sim \Lambda^{-\frac{1}{2}}_4\sim N^{\frac{1}{4}(f_1+f_2+f_3)}$ and the energy density $V^S = M_{\text{Pl}}^2\Lambda_4\sim N^{\frac{1}{2}(f_1+f_2+f_3)}$.
For completeness, we write down the KK masses in string frame $m_{\text{KK}_i}^2\sim 1/\upsilon_i\sim N^{\frac{1}{2}(f_i-f_j-f_k)}$ and as expected the KK scale is given by $L_{\text{KK}_i}^2\sim \upsilon_i\sim N^{\frac{1}{2}(-f_i+f_j+f_k)}$.
The scale separation condition becomes $L_{\text{KK}_i}^2/L_{\text{AdS}}^2\sim N^{-f_i}$.}, $M^2_{\text{Pl}}\sim N^0$. Finally, we are able to calculate how the scale separation condition depends on the general scalings
\begin{align}
    \frac{L^2_{\text{KK}_i}}{L^2_{\text{AdS}}}\equiv\frac{\langle V\rangle}{m^2_{\text{KK}_i}}\sim\frac{N^{-\frac{3}{2}(f_i+f_j+f_k)}}{N^{-\frac{1}{2}f_i-\frac{3}{2}(f_j+f_k)}}\sim N^{-f_i} \,.
\end{align}
From the above condition, we conclude that creating an anisotropy in the scaling of the fluxes and consequently in the subvolumes, requires the study of the individual decoupling of each KK scale from the AdS radius
\begin{equation}
    \frac{L_{\text{KK}_1}^2}{L_{\text{AdS}}^2}\sim N^{-f_1}\,,\quad\quad
    \frac{L_{\text{KK}_2}^2}{L_{\text{AdS}}^2}\sim N^{-f_2}\,,\quad\quad
    \frac{L_{\text{KK}_3}^2}{L_{\text{AdS}}^2}\sim N^{-f_3} \,.
\end{equation}
Since $N$ is unbounded and can be taken to be infinitely large to achieve scale separation, the scaling (exponent coefficient) of the scale separation condition should be a positive number, $f_i>0$.
Scale separation can be broken when at least one of the $f_i$ fluxes is zero, which we will refer to as partially broken scale separation. In any case, when at least one of the KK scales becomes comparable to the AdS scale, $L_{\text{KK}_i}^2\sim L_{\text{AdS}}^2$, the effective theory is affected by the presence of extra-dimensional modes. Also, it is important to notice that scale separation can never be completely broken, it can only be partially broken otherwise the overall internal space volume becomes zero or negative when $\sum_{i=1}^3f_i \geq 0$, so at least one of the $f_i$ should be positive.

For isotropic flux scaling, $f_1=f_2=f_3=1$, we recover the DGKT model, and the only free parameter is $N$.
The isotropic flux scaling is an interesting ansatz that leads to vacua with properties such as large volume, weak coupling and scale separation for large values of $N$, and it has the following scaling
\begin{align}
    \upsilon_i\sim N^{\frac{1}{2}}\,,\quad\quad e^{\phi}\sim N^{-\frac{3}{4}}\,,\quad\quad
    \text{vol}\sim N^{\frac{3}{4}}\,,\quad\quad
    \frac{L_{\text{KK}_i}^2}{L_{\text{AdS}}^2}\sim N^{-1} \,,
\end{align}
however, as we will see, anisotropic internal spaces can also have these properties.

\section{New scaling solutions}\label{new}
In this section, we present the main result of the paper by exploring various solutions to the system of equations \eqref{scaleq1}-\eqref{scaleq44} with different requirements for subvolume behavior and scale separation.
We find four new general families (regimes) of solutions while also arguing that the scaling of the original DGKT model represents a special case in the flux-scaling landscape (parameter space of scalings).

\subsection{Anisotropic scale separation, weak coupling \& large volume}\label{fam1}
In this subsection, we present the conditions that the flux scalings $f_i$ must satisfy to keep the solution within the supergravity regime while keeping the KK scale and the AdS radius parametrically separated:
\begin{align}
    \upsilon_i\gg 1\,,\quad\quad
    \text{vol}\gg 1\,,\quad\quad
    e^{\phi}< 1\,,\quad\quad
    \frac{L_{\text{KK}_i}^2}{L_{\text{AdS}}^2}\ll 1\,,
\end{align}
with $i=1,2,3$. These conditions are satisfied by DGKT, however we discover a wide range of conditions on the scaling of fluxes, and consequently, a flux-scaling landscape in which they hold. To simplify the analysis, we initially consider non-negative scalings, i.e., $f_i\geq 0$. We will later discuss the implications of allowing negative scalings.
\begin{enumerate}
\item In the first solution the scaling should satisfy the following relations
\begin{align}\label{DGKTsolution}
    f_1>0\,,\quad\quad 0<f_2\leq f_1
    \,,\quad\quad f_1-f_2<f_3<f_1+f_2 \,.
\end{align}
The DGKT vacua lie in this solution \eqref{DGKTsolution}, and its isotropic scaling $f_1=f_2=f_3=1$, satisfies the conditions due to the equality in the second relation. This can, of course, be generalized to $f_i=f_j=f_k$ with $i,j,k=1,2,3$.
We call such solutions \textit{isotropic} since the isotropy in the flux scaling imposes an isotropy to the subvolumes, or vice versa. However, there are ansatzes like $f_1=f_2=2$ and $f_3=3$ that satisfy \eqref{DGKTsolution}.
These ansatzes result in positive scaling of the subvolumes, although they do so at different rates, resulting in an \textit{anisotropy} between them: $\upsilon_1\sim N^{3/2}$, $\upsilon_2\sim N^{3/2}$, and $\upsilon_3\sim N^{1/2}$ while the relevant scales, $L_{\text{KK}_i}^2/L_{\text{AdS}}^2$, decouple as follows $N^{-1}$, $N^{-1}$ and $N^{-3/2}$.
In this family of solutions, for large $N$, we are always in the large volume regime with weak coupling, and scale separation is realized, although the rates of scaling differ.

Let us consider a type of ansatz where the behavior we want to demonstrate is more intense. For $f_1=2.8$, $f_2=1$, and $f_3=2$, the subvolumes \eqref{subvolumes} have the following scaling
\begin{align}
    \upsilon_1\sim N^{0.1}\,,\quad\quad
    \upsilon_2\sim N^{1.9}\,,\quad\quad
    \upsilon_3\sim N^{0.9}\,,
\end{align}
and from the ratio of them we observe the parametric anisotropy that is created
\begin{align}\label{parametric}
    \frac{\upsilon_1}{\upsilon_2}\sim N^{-1.8} \,,\quad\quad
    \frac{\upsilon_1}{\upsilon_3}\sim N^{-0.8} \,,\quad\quad
    \frac{\upsilon_3}{\upsilon_2}\sim N^{-1} \,.
\end{align}
while the degree of anisotropy is determined by the scaling values.
Clearly, for DGKT, the ratios would be equal to one, indicating that the radii expand isotropically for large values of $N$.

\item 
Within this category, there exists another solution where the scalings satisfy the following inequalities
\begin{align}\label{conditions1}
    f_1>0\,,\quad\quad f_2>f_1
    \,,\quad\quad -f_1+f_2<f_3<f_1+f_2 \,.
\end{align}
One can observe that, due to the lack of equality between the scalings, this solution cannot be isotropic.
\end{enumerate}

\subsection{Solutions with small subvolumes}

In this subsection we introduce solutions for scalings that possess a common feature: the presence of one or two consistently existing small subvolumes. This characteristic leads to small subvolumes \textit{shrinking} for large values of $N$, while the overall volume remains large and the $g_s$ coupling small.
Consequently, such vacuum solutions cannot be considered reliable, as the sizes for large values of $N$ become comparable to or smaller than the string length.
Nonetheless, these solutions also arise when utilizing (anti)-D4 branes to interpolate between vacua.

\subsubsection{Scale separation, weak coupling \& one small (shrinking) subvolume}\label{fam2}

We impose one subvolume to scale negatively, shrinking for large values of $N$, while the rest of the volumes and couplings become large and small respectively, while scale separation is realized
\begin{align}
    \upsilon_i< 1\,,\quad\quad
    \upsilon_j\gg 1\,,\quad\quad
    \text{vol}\gg 1\,,\quad\quad
    e^{\phi}< 1\,,\quad\quad
    \frac{L_{\text{KK}_i}^2}{L_{\text{AdS}}^2}\ll 1\,,
\end{align}
with $i=1,2,3$ and we also assume $f_i\geq 0$. For this behavior to exist, the scalings must satisfy the following conditions
\begin{equation}\label{conditions2}
f_1>0\,,\quad\quad 0<f_2<f_1\,, \quad\quad 0<f_3<f_1-f_2 \,.
\end{equation}
We obtain the same inequalities for $f_i>0$. We note that a setup with one shrinking subvolume based on a specific scaling ansatz was also recently discussed in \cite{Carrasco:2023hta}. In our examples, we will consider the simplest ansatz with small integer scalings for which the relations in \eqref{conditions2} are satisfied:
\begin{align}
    f_1=4\,,\quad\quad f_2=1 \,,\quad\quad f_3=2 \,.
\end{align}
It is important to note that the system \eqref{DGKTsolution}-\eqref{conditions1} cannot be solved when we require weak coupling, scale separation, and more than one subvolume to shrink, nor does it have a solution when we require partial or total broken scale separation.
This implies that the potential, and equivalently the equations of motion, are not invariant under a scaling that would support such behavior.

\subsubsection{Scale separation, weak coupling \& small constant subvolumes}\label{fam3}
We impose one or two of the subvolumes to remain constant for all values of $N$, while the remaining volumes and weak coupling become large and small respectively, while scale separation is realized:
\begin{align}\label{sub1}
    \upsilon_1=1\,,\quad\quad 
    \upsilon_j\gg 1\,,\quad\quad
    \text{vol}\gg 1\,,\quad\quad
    e^{\phi}< 1\,,\quad\quad
    \frac{L_{\text{KK}_i}^2}{L_{\text{AdS}}^2}\ll 1\,,
\end{align}
with $i=1,2,3$ and $j=2,3$. Since at least one subvolume is comparable to the size of the string length, we cannot trust our supergravity solution, which includes only classical ingredients. Since the fluctuations of the internal space are described by three subvolumes, analogous conditions apply to each of them, and we explicitly state them:
\begin{enumerate}
\item  The conditions imposed in \eqref{sub1}, where the first subvolume does not scale ($\upsilon_1\sim N^0$), correspond to the following solution for the scalings
\begin{equation}\label{conditions31}
    -f_1+f_2+f_3=0 \,,
    \quad\quad
    f_1>0 \,,
    \quad\quad 
    0<f_2<f_1 \,,
    \quad\quad 
    f_3=f_1-f_2 \,.
\end{equation}
\item 
Similarly, we require only the second subvolume not to scale ($\upsilon_2\sim N^0$)
\begin{equation}\label{conditions32}
    f_1-f_2+f_3=0 \,,
    \quad\quad
    f_1>0\,,
    \quad\quad f_2>f_1 \,,
    \quad\quad f_3=-f_1+f_2 \,.
\end{equation}
\item 
Lastly, we only require the third subvolume not to scale ($\upsilon_3\sim N^0$)
\begin{equation}\label{conditions33}
    f_1+f_2-f_3=0 \,,\quad\quad
    f_1>0\,,\quad\quad 
    f_2>0 \,,\quad\quad 
    f_3=f_1+f_2 \,.
\end{equation}
\end{enumerate}

\subsubsection{Broken scale separation, weak coupling \& one small subvolume}\label{fam4}

As mentioned in the previous section, our analysis shows that scale separation can be broken (or partially broken) when at least one of the scalings $f_i$ is set to zero, as we explore the scaling landscape where $f_i>0$. In this case, we partially break scale separation while remaining at weak coupling, but one subvolume still has to shrink:
\begin{align}
    \upsilon_3<1 \,,\quad\quad 
    \upsilon_j\gg 1\,,\quad\quad
    \text{vol}\gg 1\,,\quad\quad
    e^{\phi}< 1\,,\quad\quad
    \frac{L_{\text{KK}_1}^2}{L_{\text{AdS}}^2}\sim 1\,,\quad\quad
    \frac{L_{\text{KK}_j}^2}{L_{\text{AdS}}^2}\ll 1\,,
\end{align}
with $j=2,3$. There is one solution for this case with
\begin{align}\label{conditions4}
    f_1=0\,,\quad\quad f_2>0\,,\quad\quad f_3>f_2\,,
\end{align}
and the analogous ones for setting each of the other scalings to vanish. Other than these solutions, we did not find any that satisfy the requirements of broken scale separation and large/small volume. For example, when we require more than one subvolume to be small while scale separation is broken, the system does not have a solution.

\paragraph{Comments on on/off scale separation at large volume \& negative flux scaling}\

\vspace{0.1cm}

Allowing the scaling $f_i$ to take negative values expands the flux scaling landscape. It would be interesting to investigate whether one can turn scale separation on and off, similar to the study in \cite{Farakos:2023nms}, by choosing the flux scaling appropriately while staying within the supergravity limit. So far, we have found that such feature cannot be solution of the scaling equations \eqref{scaleq1}-\eqref{scaleq44}, it would be interesting to see if this can be achieved with negative values of flux scalings.

First, let us focus on the scale separation condition and argue as follows: Since its scaling depends on individual fluxes, $L_{\text{KK}_i}^2/L_{\text{AdS}}^2\sim N^{-f_i}$, scale separation would be broken for negative values of $f_i$. These solutions would all fall into a regime where at least one subvolume is small. For instance, if we assume that $f_1<0$ and $f_2>0$, $f_3>0$ to break scale separation ($L_{\text{KK}_1}^2\sim L_{\text{AdS}}^2$) while keeping one subvolume large ($\upsilon_1>1$), we would need $f_3>f_2+|f_1|$. However, to also keep the second subvolume large ($\upsilon_2>1$), we would need $f_3<f_2-|f_1|$, which is not possible due to the previous condition. Therefore, it is not possible to have broken scale separation and remain in the supergravity regime simultaneously.

\section{Interpolating between vacua}\label{section5}

In the original isotropic DGKT model, distinct values of the flux parameter $N$ correspond to different vacua, whereas scale separation is realized for large values of $N$.
In \cite{Shiu:2022oti}, the distance between non-scale-separated and scale-separated vacua was measured using space-filling D4-branes, which were codimension-one in the six directions transverse to the three four-cycles related to $F_4$ (with each cycle carrying flux $N$).
Then, by employing a parameterized modulus and allowing it to vary, D4-branes are created at a vacuum with flux $N$ and annihilate against themselves leaving behind a flux $N+1$ at the new vacuum.
This method enables us to measure the distance through a non-dynamical process.

In this section we utilize a single space-filling anti-D4-brane, denoted by $\overline{\text{D}4}$-brane, which is codimension-one in the six directions transverse to one of the three four-cycles associated to one component of $F_4$. 
Since each jump corresponds to a change in $N$, we keep $N$ constant and interpret this change as a modification to the scaling of the relevant $F_4$ flux. 
We discover that the $\overline{\text{D}4}$-brane, connects vacua with different characteristics found in subsection \ref{new}. 
 In the next section, we apply this setup to calculate the distance between vacua, where scale separation is realized within the supergravity limit. Additionally, we aim to estimate the $\gamma$ parameter of the distance conjecture \cite{Ooguri:2006in} to investigate its dependence on the change of scaling.

\subsection{Probe (anti-) D4-brane}

We follow the steps described in \cite{Shiu:2022oti} (but for an $\overline{\text{D}4}$-brane) and introduce an $\overline{\text{D}4}$-brane that spans the external space and is transverse to the four-cycle $\tilde{w}^1$, which is filled by the flux component $F_4^{(1)}$. Therefore, the brane is codimension-one in the six direction transverse to $\tilde{w}^1$. 
\begin{table}[htbp!]
\centering
\renewcommand{\arraystretch}{1.4}
\begin{tabular}{|c|c||c|c||c|c|c|c|}
\hline  & $\text{AdS}_4$ & $y^1$ & $\tilde{y}^2$ & $y^3$ & $y^4$ & $y^5$ & $y^6$ \\ 
\hline ~ $\overline{\text{D}4}$ (or $\text{D}4$)   & $\otimes$ & $\otimes$ & -- & -- & -- & -- & -- \\
\hline ~$F_{4}^{(1)}\sim N^{f_1}$ & --  & -- & -- & $\otimes$ & $\otimes$ & $\otimes$ & $\otimes$ \\
\hline ~$F_{4}^{(2)}\sim N^{f_2}$ & --  & $\otimes$ & $\otimes$ & -- & -- & $\otimes$ & $\otimes$ \\
\hline ~$F_{4}^{(3)}\sim N^{f_3}$ & --  & $\otimes$ & $\otimes$ & $\otimes$ & $\otimes$ & -- & -- \\
\hline
\end{tabular}
\caption{
The table shows the cycles filled with $F_4$ fluxes and the localized positions of the (anti)-D4-brane.
If we had chosen (anti)-D4-branes to be codimension-one in the six directions transverse to all the four cycles $\tilde{w}^i$, we would not be able to interpolate between vacua associated with the different regimes we discovered in the previous section. 
This is because the value of the relevant fluxes and the dependent quantities would change isotropically for different values of $N$, and thus we would have remained in the initial regime.}\label{D4table} 
\end{table}
For reasons that we will explain in a short, we will use an $\overline{\text{D}4}$-brane instead of a D4-brane, but the process remains the same. The Bianchi identity for an (anti)-D4-brane, filling the $\text{AdS}_4$ space and localized to the coordinates of the internal space, is given by
\begin{align}
    \text{d}F_4=\pm (2\pi)^7\alpha^{\prime 4}Q_{\text{D4}}\delta (\tilde{y}^2,y^3,y^4,y^5,y^6)\text{d}\tilde{y}^2\wedge \epsilon_4 \,,
\end{align}
where $\epsilon_4=\text{d}y^3\wedge\dots\wedge\text{d}y^6$ is the normalized volume form of the four-cycle that $F_4$ is filling, and its directions, along with $\tilde{y}^2$-direction, span the space transverse to the (anti)-D4-brane. The forms $\tilde{w}^1\sim\epsilon_4$ are equivalent up to a numerical factor. It is worth noting that for BPS objects, the tension is equal to the charge, and in the Bianchi identities, we use the D4-brane change with the opposite sign for the relevant anti-brane, i.e., $Q_{\overline{\text{D4}}}=-Q_{\text{D4}}$. 

The special position of the (anti)-D4-brane leads to the following observation: it causes a change in the $F_4$ flux on both sides of the (anti)-D4-brane, for $\tilde{y}^2<0$ and $\tilde{y}^2>0$, and can be easily seen when smearing the (anti)-D4-brane over a four-cycle $\tilde{w}^i$ since the Bianchi identity can be solved for
\begin{align}\label{jumps}
    F_4^{(i)}=\left[e_i\pm (2\pi\,l_s)^3\theta(\tilde{y}^2)\right]\epsilon_4 
    =(2\pi\,l_s)^3\left[N^{f_i}\pm \theta(\tilde{y}^2)\right]\epsilon_4\,,
\end{align}
with $i=1$, while for branes wrapping different cycles, we would have three different equations with $i=1,2,3$. The function $\theta(\tilde{y}^2)$ stands for the Heaviside function while from the quantization of the flux we used that $e_1=(2\pi\,l_s)^3\times N^{f_i}$ where $N^{f_i}$ is the relevant quantized number which can acquire large integer values $N^{f_i}\in \mathbb{Z}$.

Now we use the creation and annihilation of the (anti)-D4-brane to transition from one vacuum to another, as in \cite{Shiu:2022oti}.
The method using a varying open string modulus is described in the next section, while here we focus on the flux change.
We create such an object in the two-cycle, let it expand and contract scanning out the entire two-cycle, and annihilate against itself, leaving behind a different flux vacuum.
In our current setup, as indicated by \eqref{jumps}, each time we employ this process to transition from one vacuum to another, the fluxes undergo the following change:
\begin{align}\label{Fluxchange1}
    F_4^{(1)}=N^{f_1}\pm 1\,,\quad\quad F_4^{(2)}=N^{f_2} \,,\quad\quad F_4^{(3)}=N^{f_3} \,.
\end{align}
Since the number $N$ is not constrained by the equations of motion, it can be chosen to be large, specifically, $N\gg 1$. Consequently, each time a jump is performed, we interpret the flux change as a change in the scaling of the flux $F_4^{(1)}$ as follows:
\begin{align}\label{Fluxchange2}
    N^{f_1}\pm 1= N^{\tilde{f}_1} \,,
\end{align}
where $\tilde{f}_1$ is the scaling of the $F_4^{(1)}$ flux at the new vacuum, maintaining the same value for $N$.


\subsection{Interpolating between different regimes with anti-D4-brane}\label{inter}

From now on, we will use an $\overline{\text{D4}}$-brane to interpolate between vacua. To understand how the subvolumes and scale separation change with each jump of the anti-D4-brane, we study a simple example with a small value of $N$, more specifically for $N=2$, and then discuss how it works for large values of $N$.
As we will demonstrate, there exist vacua where the supergravity approximation is valid and scale separation is realized, surrounded by solutions where subvolumes are small, and the supergravity solution cannot be trusted.

We assume that we start from a vacuum with the same conditions as in \eqref{fam2}, where the fluxes of $F_4^{(i)}$ scale like $f_1=4$, $f_2=1$, and $f_3=2$. Every time the $\overline{\text{D}4}$-brane creates and annihilates against itself, it leaves behind a \textit{negative} flux, decreasing the previous $F_4^{(1)}$ value. We keep the value of $N$ constant and interpret the decrease in flux value as a decrease in the scaling $f_1$. This behavior is shown in Table \ref{CCtable2}. A first observation is that the scaling of the relevant scale separation condition becomes more and more positive and after of number of steps scale separation will break.

Now we turn to the subvolumes: As the scaling of the subvolume $\upsilon_1$ becomes more positive, the scalings of $\upsilon_2$ and $\upsilon_3$ decrease. After a few steps, the scaling of $\upsilon_1$ reaches zero, and the subvolume becomes constant and comparable to the string length. This behavior, where scale separation is realized, and one subvolume becomes constant, is a solution of the scaling equations described in \eqref{fam3}. However, such a solution cannot be trusted. Subsequently, as the scaling $f_1$ increases, the scaling of $\upsilon_1$ turns from zero to positive. In this regime, and after a few more jumps of the $\overline{\text{D}4}$-brane, all the volumes scale positively, while $L^{2}_{\text{KK}_i}/L_{\text{AdS}}^2$ scales negatively (scale separation is realized in the supergravity limit). This regime is described in \eqref{fam1}. For very large values of $N$, these scalings would lead to a regime with large volumes, weak coupling, and scale separation.
\begin{table}
\centering
\renewcommand{\arraystretch}{1.4}
\begin{tabular}{|c||c||c|c|c||c|c|c|c|c|}
     \hline 
     $F_4^{(1)}\sim N^4$ & $f_1$  & $\upsilon_1$ & $\upsilon_2$ & $\upsilon_3$  & $L_{\text{KK}_1}/L_{\text{AdS}}$ &  \text{Solution} &  \\ 
     \hline \hline
     $N^4=2^{f_1}$ & $4$ & $N^{-\frac{1}{2}}$  &  $N^{\frac{5}{2}}$ & $N^{\frac{3}{2}}$ & $N^{-2}$ & \eqref{conditions2} & $X$  \\
     \hline
     $N^4-1=2^{f_1}$ & 3.9 & $N^{-0.54}$  &  $N^{2.45}$ &  $N^{1.45}$ & $N^{-1.8}$ & \eqref{conditions2} & $X$ \\
     \hline
     $N^4-\,$\dots$\, =2^{f_1}$ & $\dots$ & $\dots$ &  $\dots$ &  $\dots$ & $ N^{\dots}$ & \eqref{conditions2} & $X$ \\
     \hline
     $N^4-8=2^{f_1}$ & 3 & $N^{0}$ & $N^{2}$ &  $N^{1}$ & $N^{-1.5}$ & \eqref{conditions31} & $X$ \\
     \hline
     $N^4-9=2^{f_1}$ & 2.8 & $N^{0.1}$ &
     $N^{1.9}$ & $N^{0.9}$ &
     $N^{-1.4}$ & \eqref{conditions1} & $\checkmark$  \\
     \hline
     $N^4-10=2^{f_1}$ & 2.6 & $N^{0.2}$ &
     $N^{1.8}$ & $N^{0.8}$ &
     $N^{-1.3}$ & \eqref{conditions1} &  $\checkmark$ \\
     \hline
     $N^4-11=2^{f_1}$ & 2.3 & $N^{0.34}$ & $N^{1.66}$ & $N^{0.66}$ & $N^{-1.15}$ & \eqref{conditions1} &  $\checkmark$ \\
     \hline
     $N^4-12=2^{f_1}$ & 2 & $N^{1/2}$ & $N^{3/2}$ & $ N^{1/2}$ & $N^{-1}$ & \eqref{conditions1} & $\checkmark$ \\
     \hline
     $N^4-13=2^{f_1}$ & 1.6 & $N^{0.7}$  & $N^{1.3}$ & $N^{0.3}$ & $N^{-0.8}$ & \eqref{conditions1} &  $\checkmark$ \\
     \hline
     $N^4-14=2^{f_1}$ & 1 & $N^{1}$  &  $N^{1}$ &  $N^{0}$ & $N^{-0.5}$ & \eqref{conditions33} &  $X$ \\
     \hline
     $N^4-15=2^{f_1}$ & 0 & $N^{3/2}$  &  $N^{1/2}$ & $N^{-1/2}$ & $N^{0}$ & \eqref{conditions4} & $X$  \\
     \hline
\end{tabular}
\caption{\label{CCtable2}
The table presents the changes in the scaling of $F_4^{(1)}$ flux, subvolumes, and scale separation when the $\overline{\text{D}4}$-brane interpolates between vacua and different regimes presented in section \ref{new}.
The checkmark stands for the vacuum solutions where the supergravity approximation is valid and scale separation is realized.
It can be seen that it takes 9 jumps for the scaling of $\upsilon_1$ to become positive (supergravity regime for large $N$) and after 5 jumps the scaling of $\upsilon_3$ to becomes zero (would exist the supergravity regime for large $N$). One could have used a D4-brane instead, however, the creation and annihilation of the D4-brane would increase the flux by $N^{f_1}+1$, changing the $f_1$ scaling positively. This would require starting from the vacuum with $f_1=0$ to interpolate through all regimes.}
\end{table}
\paragraph{Generalization for large $\textit{\textbf{N}}$} 
To generalize, we keep the same values for the flux scalings but change $N$ to $N=3$. We find that it takes 54 jumps, starting from the initial vacuum with $f_1=4$, for the subvolume $\upsilon_1$ to become positive and this happens when $f_1=3$. This is reasonable as the change in the sign of the scaling of $\upsilon_1$ depends only on the initial values of the other two flux scalings which remain unaffected by the jumps.  

Let us conclude: for any value of $N$ and for a subvolume $\upsilon_i$, only the initial conditions of the flux scalings $f_j$ and $f_k$ determine whether we are in the supergravity regime, as the scaling of subvolumes will change sign when $f_i=f_j+f_k$. It is also obvious that scale separation breaks when $f_i$ approaches zero. The value of $N$ just determines the number of steps required to reach and interpolate through such regimes, with the number of steps being $N^{f_1^{initial}}-N^{f_1^{final}}$. This implies that $N$ can be chosen to be arbitrarily large since the control of the supergravity regime relies on the flux scalings.

It is important for the calculations in the next section to consider how $f_1$ changes with increasing values of $N$.
In table \ref{CCtable2}, we observe that the value of the $f_1$ scaling is approximately $f_1 \approx 1.6$ before $\upsilon_3$ becomes constant.
For large values of $N$, one step before $\upsilon_3$ becomes constant, the value of $f_1$ is approximately the same as that of $\upsilon_3$ when it becomes constant.

\section{Calculation of the distance for the classical supergravity regime}

As previously discussed, the non-dynamical process of creating and annihilating the (anti)-D4-brane involves using an open string modulus that describes its localized position within a two-cycle. 
In this section, as an application, we introduce and allow the modulus to vary and calculate the distance between vacua where scale separation is realized within the supergravity limit. We also estimate the $\gamma$ parameter of the distance conjecture \cite{Ooguri:2006in}.
Then, we argue that the distance between vacua with different scalings as well as the parameter $\gamma$ both depend on the values of the scaling.
For further discussions on geodesic distance and discrete vacua using domain walls, see \cite{Basile:2023rvm}.
For details on the explicit derivation of the (anti)-D4-brane action, we refer to \cite{Shiu:2022oti} and \cite{Farakos:2023nms}. In this section we will use the conventions (coupling, frames, units) presented in the former.

\subsection{EFT of the anti-D4-brane and open string modulus}
We use the notation of \cite{Shiu:2022oti} and the local description of the metric is given by:
\begin{align}
    \text{d}s^2_{10}=g_{\mu\nu}\text{d}x^{\mu}\text{d}x^{\nu}+l^2_s\upsilon_1\left[f^2(\psi)\text{d}\phi^2+\text{d}\psi^2\right]+\text{d}s^2_{\Sigma_4} \,.
\end{align}
For our purposes we have expressed the relevant two-cycles in some general spherical coordinates while $\text{d}s^2_{\Sigma_4}$ represents the space filled with $F_4^{(1)}$ flux. The local object fills the AdS$_4$ space, wraps a circular direction $\phi$, while $\psi$ is a localized position which parametrizes the movement of the $\overline{\text{D}4}$-brane inside the two-cycle. This scanning of the two-cycle will be performed by parametrizing a modulus $\psi$ and allowing it to vary. For example, for a two-sphere the function $f(\psi) = \text{sin}\,\psi$ allows us to visualize the creation and annihilation of the $\overline{\text{D}4}$-brane and the vacuum interpolation. Imagine that there is vacuum characterized by flux $N^{f_1}$ and the field has the value $\psi = 0$, which corresponds to the one pole of the two-sphere, then we force it to move to the point $\psi = \pi$, which corresponds to the other pole, where it annihilates, leaving behind a flux with $N^{f_1}-1=N^{\tilde{f}_1}$.

Now we let $\psi$ vary with spacetime coordinates, perform the pullback\footnote{The pulled-back metric one the brane is $\text{d}s^2_{\text{D4}}=\left(g_{\mu\nu}+l_s^2\upsilon_1\partial_{\mu}\psi\partial_{\nu}\psi\right)\text{d}x^{\mu}\text{d}x^{\nu}+l_s^2\upsilon_1f^2(\psi)\text{d}\varphi^2$.} and expand the DBI part of the $\overline{\text{D}4}$-brane action, we obtain:
\begin{align}
    S_{\text{D4}}=-2\pi l_s\mu_{\text{D}4}\int \text{d}^4x\sqrt{-g_4}\left[\frac{1}{2}l_s^2\upsilon^{\frac{3}{2}}_1e^{-\phi}\vert f(\psi)\vert\partial_{\mu}\psi\partial^{\mu}\psi+\sqrt{\upsilon_1}e^{-\phi}\vert f(\psi)\vert \right] \,.
\end{align}
From there it is straightforward to read off the $\psi$-field moduli space metric $g_{\psi\psi}$ and the potential $V(\psi)$ which we won't use it in this analysis. Using the relation $8\pi^2\gamma M^2_{\text{Pl}}=\text{vol} \,e^{-2\phi}l_s^{-2}$, where $\gamma$ is an order-one constant, the relevant field metric takes the following form 
\begin{align}
    \frac{g_{\psi\psi}}{M^2_{\text{Pl}}}=\gamma\frac{\sqrt{\upsilon_1}}{\upsilon_2\upsilon_3}e^{\phi}\vert f(\psi)\vert \,.
\end{align}

\subsection{Geodesic distance and distance conjecture}
In this subsection, we calculate the distance between two points in moduli space that correspond to the boundaries of a region where scale separation is realized, and classical supergravity is valid. Outside this region, the subvolumes either remain constant or shrink. Concerning our setup, the $\overline{\text{D4}}$-brane wrapping a single direction in the internal space reminds us of \cite{Farakos:2023nms}, but the field content is entirely different. The field content resembles that of the isotropic case in \cite{Shiu:2022oti}, although now the $T^2$ deformations are anisotropic and treated as distinct moduli.

For a theory with scalar field $\varphi^A$, the distance $\Delta$ is given by
\begin{align}
\Delta=\int_{\xi=0}^{\xi=1}\text{d}\xi\sqrt{\frac{g_{AB}}{M_{\text{Pl}}^2}\frac{\text{d}\varphi^A}{\text{d}\xi}\frac{\text{d}\varphi^B}{\text{d}\xi}}
\end{align}
where $g_{AB}$ stands the moduli metric. We introduce the four-dimensional dilaton
\begin{align}
    e^D=\frac{e^{\phi}}{\sqrt{\text{vol}}}\sim N^{-\frac{1}{2}(f_1+f_2+f_3)} \,,
\end{align}
and we write down our moduli as $\upsilon_i\sim e^{2\sigma_i}$ such that the 4d dilaton is $D=\phi-\sum_{i=1}^3\sigma_i$ with the string-frame metric to be
\begin{align}
    \text{d}s^2_{10}=e^{2D}\text{d}x^2_4+\sum_{i=1}^3e^{2\sigma_i}\text{d}y^2_{2,i} \,.
\end{align}
The field content is now consisted by the following moduli
\begin{align}
    \varphi^A: \{D,\psi,\sigma_1,\sigma_2,\sigma_3\} \,,
\end{align}
and write down the remaining metric components that we will use for the calculation of the distance $g_{DD}= 2M^2_{\text{Pl}}$ and $g_{\sigma_i\sigma_i}= 2M^2_{\text{Pl}}$. Then the distance can be written as
\begin{align}
    \Delta
    &=\int^1_0\text{d}s\sqrt{\gamma\,e^{D+\sum_i\sigma_i}e^{\sigma_1-2\sigma_2-2\sigma_3}\vert f(\psi)\vert \left(\frac{\text{d}\psi}{\text{d}s}\right)^2
    +2\left(\frac{\text{d}D}{\text{d}s}\right)^2
    +2\sum_{i=1}^3\left(\frac{\text{d}\sigma_i}{\text{d}s}\right)^2} \,,
\end{align}
and perform the following redefinition
\begin{align}
    \frac{\text{d}\chi}{\text{d}\psi}=\sqrt{\frac{\gamma}{8}}\sqrt{\vert f(\psi)\vert} \,,
\end{align}
such that the geodesic distance for the moduli space takes the form
\begin{align}
    \Delta=\int^1_0\text{d}s\sqrt{8\,e^{D+2\sigma_1-\sigma_2-\sigma_3} \left(\frac{\text{d}\chi}{\text{d}s}\right)^2
    +2\left(\frac{\text{d}D}{\text{d}s}\right)^2
    +2\sum_{i=1}^3\left(\frac{\text{d}\sigma_i}{\text{d}s}\right)^2} \,.
\end{align}
To proceed, we introduce the new coordinates $\{\tilde{D},\tilde{\sigma}_1,\tilde{\sigma}_2,\tilde{\sigma}_3\}$ by performing an O(4) rotation\footnote{The O(4) rotation is given by
\begin{equation}\label{rotation}
\left(\begin{matrix} D \\ \sigma_1 \\ \sigma_2 \\ \sigma_3 \end{matrix}\right) = \frac{1}{\sqrt{\text{det}\,O}} \left(\begin{matrix}  a & -b & -c & -d
\\ b & a & -d & c
\\ c & d & a & -b
\\ d & -c & b & a \end{matrix}\right)  \left(\begin{matrix} \tilde{D} \\ \tilde{\sigma}_1 \\ \tilde{\sigma}_2 \\ \tilde{\sigma}_3 \, \end{matrix}\right), \quad\quad \text{det}\,O=(a^2+b^2+c^2+d^2)^2
\end{equation}
where $O$ stands for the $4\times 4$ matrix and in our case $a=1\,,b=2\,,c=-1\,,d=-1$ and $\text{det}\, O=49.$}
and redefine the angular variable  
\begin{align}
\chi = \frac{1}{(\text{det}\,O)^{\frac14}} h_1 \,,
\end{align}
such that the geodesic distance for the moduli space distance takes the form
\begin{align}
\Delta = \frac{1}{(\text{det}\,O)^{\frac14}} \int_0^1 \text{d}\xi \sqrt{8e^{\tilde{D}} \left(\frac{\text{d}h_1}{d\xi}\right)^2 
+2\left(\frac{\text{d}\tilde{D}}{d\xi}\right)^2 +2\left(\frac{\text{d}\tilde{\sigma}_1}{d\xi}\right)^2 
+2\left(\frac{\text{d}\tilde{\sigma}_2}{d\xi}\right)^2 
+2\left(\frac{\text{d}\tilde{\sigma}_3}{d\xi}\right)^2} \,.
\end{align}
Finally after trading $\tilde{D}$ for $h_2 = e^{-\tilde{D}/2}$ the geodesic distance becomes
\begin{align}\label{Geodesic1}
\Delta = \frac{1}{(\text{det}\,O)^{\frac14}} \int_0^1 \text{d}\xi \sqrt{\frac{8}{h_2^2} \left[\left(\frac{\text{d}h_1}{d\xi}\right)^2 +\left(\frac{\text{d}h_2}{d\xi}\right)^2 \right] 
+2\left(\frac{\text{d}\tilde{\sigma}_1}{d\xi}\right)^2 
+2\left(\frac{\text{d}\tilde{\sigma}_2}{d\xi}\right)^2 
+2\left(\frac{\text{d}\tilde{\sigma}_3}{d\xi}\right)^2}
\end{align}
which corresponds to a $\mathbb{H}^2 \times \mathbb{R}^3$ space as in \cite{Farakos:2023nms}. The geodesic path is given then by \cite{Shiu:2022oti},\cite{Farakos:2023nms}
\begin{align}\label{system}
h_1(\xi)= l_0 \frac{e^{2(d_1 \xi + d_2)}-1}{e^{2(d_1 \xi + d_2)}+1} + h_{0} \,,\quad\quad 
h_2(\xi)= 2 l_0 \frac{e^{d_1 \xi + d_2}}{e^{2(d_1 \xi + d_2)}+1}
\end{align}
and
\begin{align}\label{sigmas}
\tilde{\sigma}_1(\xi) = d_3 \xi + d_4 \,, \quad\quad \tilde{\sigma}_2(\xi) = d_5 \xi + d_6 \,, \quad\quad \tilde{\sigma}_3(\xi) = d_7 \xi + d_8 \,,
\end{align}
where $d_i$, $l_0$ and $h_0$ are integration constants, and considering the above, the geodesic distance evaluates to 
\begin{align}\label{Geodesic2}
\Delta = \frac{1}{(\text{det}\,O)^{\frac14}}  \sqrt{8d_1^2 + 2d_3^2 + 2d_5^2 + 2d_7^2} \,.
\end{align}
To calculate the distance, we must determine the coefficients ${d_i}_{i=1,3,5,7}$. This involves calculating the values of the redefined fields, $h_1({\xi})$, $h_2({\xi})$, $\tilde{\sigma}_2({\xi})$, $\tilde{\sigma}_3({\xi})$, and $\tilde{\sigma}_4({\xi})$, for the two vacua between which we want to interpolate. To ensure the validity of our classical solution, we focus on the region in the moduli space where all the subvolumes are large (and scale separation is realized). In this context, $\xi=0$ corresponds to the vacuum where $f_1\approx 3$, while $\xi=1$ corresponds to $\tilde{f}_1\approx 1$ (For details on those scaling values, see the discussion in the last paragraph of subsection \ref{inter} and Table \ref{CCtable2}). It is important to remind to the reader that in our case, $f_1$ changes due to the jumps of the $\overline{\text{D4}}$-brane, whereas $f_2=1$ and $f_3=2$ remain constant. The coefficients $d_3, d_5, d_7$ can be determined from \eqref{sigmas} at $\xi=1$
\begin{align}
    d_3\sim \text{log}\,N^{\frac{1}{4}(-\tilde{f}_1+f_2+f_3)} \,,\quad\quad 
    d_5\sim \text{log}\,N^{\frac{1}{4}(\tilde{f}_1-f_2+f_3)} \,,\quad\quad
    d_7\sim \text{log}\,N^{\frac{1}{4}(\tilde{f}_1+f_2-f_3)} \,,
\end{align}
and considering the scaling values at this vacuum, we find that $d_3\sim\text{log}\,N^{1/2}$, $d_4\sim\text{log}\,N^{1/2}$, and $d_5\sim\text{log}\,N^0$. To estimate $d_1$, we need to solve the system of equations in \eqref{system} at the boundaries. The redefined fields have the following scalings: $h_1(0)\sim 1$ and $h_1(1)\sim N$, while $h_2(0)\sim 1$ and $h_2(1)=e^{-D/2}\sim N^{\frac{1}{4}(\tilde{f}_1+f_2+f_3)}$. For the calculation of the distance, we only need to determine $d_1$ and taking all of the above into account we find that
\begin{align}
    d_1=\text{log}\left[\frac{1}{2}N^{-a/4}\left(1+N^2+N^{a/2}+\sqrt{4N^2+(-1+N^2+N^{a/2})^2}\right)\right],
\end{align}
with $a=f_1+f_2+f_3$. We see that for large values of $N$, the leading-order term of $d_1$ depends on the values of the exponents\footnote{For $a=3$, which corresponds to the isotropic DGKT discussed in \cite{Shiu:2022oti}, we restore their scaling as $d_1\sim N^{5/4}$.} since in our example $f_1$ changes, the sum of the exponents changes in each step. To proceed, we will not keep the scaling general but focus on our specific case of interest. We are interested in the position $\xi=1$ where, for large values of $N$, $f_1\equiv\tilde{f}_1\approx 1$ and thus $a\approx 4$. Therefore, for large $N$, we find that $d_1\sim\text{log}\,N$, and using \eqref{Geodesic2}, we obtain
\begin{equation}\label{del}
    \Delta\sim \frac{3}{\sqrt{7}}\text{log}\,N \,.
\end{equation}
We refer to the formula for the distance conjecture to calculate the $\gamma$ coefficient
\begin{align}\label{gam}
    m^2_{\text{KK}}\vert_{\tilde{f}_1}= m^2_{\text{KK}}\vert_{f_1}\,e^{-\gamma \Delta} 
    &\,\rightarrow\, \frac{1}{2}(f_1-\tilde{f}_1)\text{log}\,N= \gamma \Delta \,,
\end{align}
where we have not replaced the scaling values to emphasize their dependence on the relation. The scaling dependence arises from the KK masses in the string frame, and finally, by comparing \eqref{del} and \eqref{gam}, we find that
\begin{align}
    \gamma\sim\frac{\sqrt{7}}{6}(f_1-\tilde{f}_1)\sim 0.88 \,.
\end{align}
It is important to clarify once again that we measured the distance between two points in the moduli space, both of which correspond to scale-separated vacua where the supergravity approximation is valid. 
It is also important to clarify the following: one should not assume that choosing arbitrarily large values for the scalings (in line with the conditions in section \ref{new}) will increase the distance and consequently change the value of $\gamma$. This is because the swampland distance conjecture depends also on the geodesic distance $\Delta$, which will also change when changing the values of the scaling. However, we have determined that the $\gamma$ parameter is independent of the flux $N$, and we have identified its scaling dependence.


\section{Conclusions} 
In this paper we investigated supersymmetric AdS$_4$ vacua resulting from the compactification of massive type IIA on $T^6/\mathbb{Z}_3\times \mathbb{Z}_3$, where $T^6$ is a product of three two-tori $T^2\times T^2\times T^2$.
Our main results arise from a general scaling analysis that reveals four distinct regimes with different characteristics related to scale separation and volume sizes, determined by the different scaling of the components of the $F_4$ flux. 
We highlight the significant role played by different scalings of the $F_4$ flux, leading to anisotropic internal space and classical solutions with scale separation.
The anisotropic internal space is a result of the different scaling of the subvolumes, and the degree of anisotropy is determined by the scaling values, which are dictated by inequalities.
We then employed a novel approach involving an anti-D4-brane, which enables us to interpolate between classical supergravity solutions with scale separation, surrounded by vacua characterized by small subvolumes compared to the string length.
Furthermore, we explored the distances within this landscape and revealed the dependence of geodesic paths and distance conjecture on anisotropic scaling.

An important question to consider in the future is understanding the precise connection between the swampland distance conjecture and general anisotropic fluxes. This can be challenging because the scalings involved in the swampland distance conjecture have complicated dependencies. It is also interesting to investigate whether flux scalings and the distance conjecture can place restrictions on each other. Additionally, exploring anisotropic behaviors and more general approaches in other models is another interesting path for future research.

\section*{Acknowledgements} 

We would like to thank David Andriot, Niccol\`o Cribiori, Fotis Farakos and Thomas Van Riet for useful comments on the draft. We would like especially to thank Vincent Van Hemelryck for valuable discussions and correspondence.


\end{document}